\documentclass[final,a4paper,onecolumn,3p,12pt]{elsarticle}

\usepackage{graphicx,amssymb,mathrsfs,amsmath}
\usepackage{bm}

\begin{document}
\title{Understanding  $Y(4274)$ and $X(4320)$ in the  $J/\psi \phi$ invariant mass spectrum}

\author[a,b,c]{Jun He}
\author[a,b,d]{Pei-Liang L\"u}

\address[a]{Nuclear Theory Group, Institute of Modern Physics, Chinese Academy of Sciences,
Lanzhou 730000, China}
\address[b]{Research Center for Hadron and CSR Physics,
Lanzhou University and Institute of Modern Physics of CAS, Lanzhou 730000, China}
\address[c]{State Key Laboratory of Theoretical Physics, Institute of
Theoretical Physics, Chinese Academy of Sciences, Beijing  100190,China}
\address[d]{University of Chinese Academy of Sciences, Beijing 100049, China}

\begin{abstract}

In this work we study the structures near 4.3 GeV in the $J/\psi \phi$
invariant mass spectra in $B$ meson decay process $B^+\to J/\psi \phi
K^+$ and two photon fusion process $\gamma\gamma\to J/\psi \phi$. The
$Y(4274)$ as a $D_sD_{s0}$ $(2317)$ molecular is studied in the Bethe-Salpeter equation
approach with quasipotential approximation. The absence of $Y(4274)$ in $\gamma\gamma\to J/\psi \phi$
channel can be well explained by the decay widths of $Y(4274)$
decaying to $\gamma\gamma$ and $J/\psi\phi$.  The distribution of mass
difference released by CMS collaboration is reproduced by two
resonances near 4.3~GeV, $Y(4274)$ and $X(4320)$.  The
different production mechanism suggests $X(4320)$ observed in the $B$
decay should be the missing $3^3P_1$ charmonium state $\chi''_{c1}$
and different from $X(4350)$ observed in two photon fusion which can
be assigned as $\chi''_{c2}$.

\end{abstract}


\maketitle
\flushbottom

\section{Introduction}

\begin{figure}[h!]
\begin{center}
\includegraphics[bb=30 50 750 720 ,scale=0.53, clip]{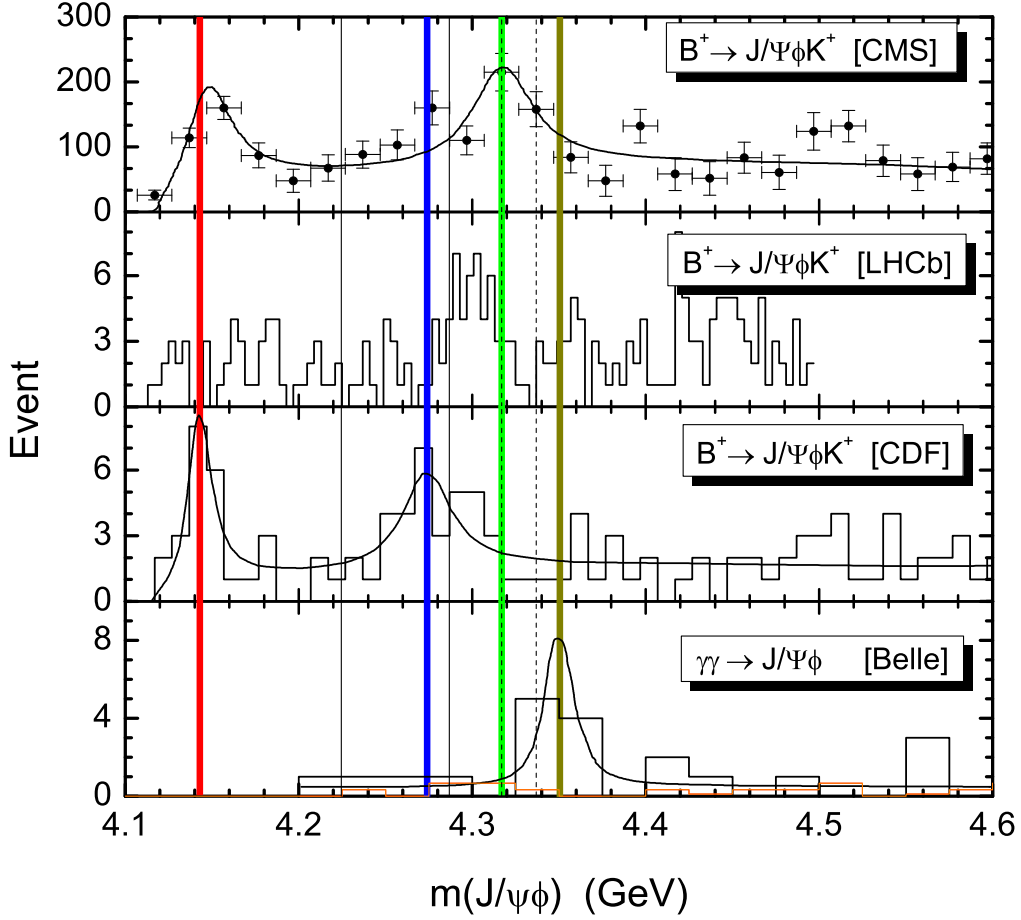}
\end{center}
\caption{The invariant mass spectrum $m(J/\psi\phi)$
by CMS collaboration~\cite{ATLAS:2013eta}, LHCb
collaboration~\cite{Aaij:2012pz}, CDF collaboration~\cite{Yi:2010aa,Aaltonen:2009tz}
and Belle collaboration~\cite{Shen:2009vs}. The three full thick vertical lines from left to right (red, blue, green and brown)
are for $Y(4140)$, $Y(4274)$, $X(4320)$ and $X(4350)$, respectively. The
curves are the fitting results by the experimental
collaborations. The two full thin vertical
lines are the thresholds for $D^*\bar{D}^*$ and
$D^*\bar{D}_{s0}(2317)$~\cite{PDG}. The two dashed thin vertical lines are the
masses of $\chi''_{c1}$ and $\chi''_{c2}$
predicted by GI potential~\cite{Barnes:2005pb}.
\label{Fig: Exp}}
\end{figure}

Very recently, CMS collaboration released their results about the
$J/\psi \phi$ spectrum in  $B^+\to J/\psi \phi
K^+$~\cite{ATLAS:2013eta}. Two peaks at mass values of
4148$\pm$2.0(stat)$\pm$4.6(syst) MeV and
4316.7$\pm$3.0(stat)$\pm$7.3(syst) MeV, remarked as X(4320) here and
hereafter, were reported. If we recall the previous experimental
results as shown in Fig.~\ref{Fig: Exp}, $J/\psi \phi$ invariant mass
spectrum near 4.3 GeV is confusing to some extent. In the energy
region near 4.3~GeV, three structures, $Y(4274)$, $X(4320)$ and
$X(4350)$ were reported by CDF Collaboration, CMS collaboration and
Belle collaboration, respectively. No evidence of $Y(4274)$ was
reported in the measurement by LHCb collaboration~\cite{Aaij:2012pz}.
However, As shown in Fig.~\ref{Fig: Exp}, there exists an obvious bump
structure near $4.3$~GeV, which was also suggested by Yi in a recent review
article~\cite{YI:2013vba}.

To understand the $J/\psi \phi$  invariant mass spectrum near 4.3 GeV, the following questions should be answered,
\begin{itemize}
  \item The structure $Y(4274)$ is found by CDF Collaboration in the $B$ decay process.  Why is it not found in two photon fusion?
  \item If $Y(4274)$ exists, it should be explained why the $Y(4274)$ is not reported by CMS collaboration in the same channel $B^+\to
J/\psi \phi K^+$.
  \item Is the structure $X(4320)$ different from $Y(4274)$ considered the large mass difference?  Why is it not found in CDF experiment?
  \item Are the structure $X(4320)$ found in the $B$ decay and the structure $X(4350)$ found in the two photon fusion the same?
\end{itemize}

Among all questions, we should first confirm the existence and the
internal structure of $Y(4274)$.  The first choice is putting
$Y(4274)$ in the frame of the constituent quark model with structure
$c\bar{c}$.  By checking the Table I in Ref.~\cite{Barnes:2005pb} and
considering $Y(4274)$ observed in the $J/\psi\phi$ invariant mass
spectrum, we can conclude that $Y(4274)$ should be a P-wave state with
the second radial excitation with the quantum number $I^G(J^{PC})=
0^+(J^{++})$ with $J = 0, 1$ or $2$. However, the calculation in
$^3P_0$ mode shows that the total widths of the
second radial excitations of $\chi_{c0}$ and $\chi_{c1}$ are larger
than the experimental observed width of $Y(4274)$ and the candidate of
the second radial excitation of $\chi_{c2}$ has be assigned to
$X(4350)$~\cite{Liu:2009fe}. Besides, because mass of
$Y(4274)$ is far beyond the open charm threshold, It would be expected
that the dominant decay channel of $Y(4274)$ is  open charm pairs and
the branching fraction in  $J/\psi\phi$ channel is
tiny~\cite{YI:2013vba}.  Hence, the assignment of $Y(4274)$ as a
charmonium is not preferred.  Of course, at present we cannot fully
exclude the P-wave charmonium explanation of Y(4274), since the
uncertainty of $^3P_0$ model is not under control.

A state with both hidden charm and hidden strange should be easy to be
produced in the $J/\psi \phi$ channel and the threshold of two
charm-strange mesons $D_s\bar{D}_{s0}(2317)$ is 4.287~GeV.
In~\cite{Shen:2010ky}, a systemic study of $D_s\bar{D}_{s0}(2317)$
in one-boson-exchange (OBE) model have been done. It has been
suggested in the literatures with one-boson-exchange (OBE) model and QCD sum rule that the structure $Y(4274)$ can be assigned
as a S-wave $D_s\bar{D}_{s0}(2317)$ molecular state with bound energy
about 10~MeV with a internal structure $(|D_s^{+}D_{s0}^{-}\rangle
+|D_s^{-}D_{s0}^{+}\rangle)/\sqrt{2}$, which has quantum number
$I^G(J^{PC})= 0^+(0^{-+})$
~\cite{Liu:2010hf,He:2011ed,Finazzo:2011he}.  In the previous
works~\cite{Liu:2010hf,He:2011ed} the mass of $Y(4274)$ have been
reproduced in a non-relativistic OBE model by solving Sch\"odinger equation. The three body decay of $Y(4274)$ is also
discussed in Ref~\cite{He:2011ed}. However, there does not exist a
theoretical study about the decays of $Y(4274)$ into $J/\psi\phi$ and $\gamma\gamma$
where it is observed.

The molecular state is a loose bound state of two hadrons, so the
Bethe-Salpeter (BS) equation is an appropriate tool to deal with the
molecular state. In Refs. \cite{Guo:2007mm,Ke:2012gm} the $K\bar{K}$
and $Z_b(10610)$ have been studied in the BS equation approach with
quasipotential approximation. Moreover, we have studied the $Y(4274)$
and its three body decay in the BS equation approach with
non-relativistic approximation \cite{He:2011ed}. And this method is
successfully applied to the $D^*_0(2400)N$ system and found
$\Sigma_c(3250)$ reported by BarBar collaboration recently can be
explained as a $D^*_0(2400)N$ molecular state \cite{He:2012zd}. To
explore the internal structure of $Y(4274)$, in this work we will
study $Y(4274)$ as $D_s\bar{D}_{s0}(2317)$ molecular state in BS
equation approach and discuss its decay pattern to explain the absence
of $Y(4274)$ in the two photon fusion.

The coincidence of the mass of $\chi''_{c2}$ predicted by
Godfrey-Isgure (GI) potential~\cite{Barnes:2005pb} and the $X(4320)$
reported by CMS collaboration as shown in Fig.~\ref{Fig: Exp} suggests
that the $X(4320)$  may be a good candidate of the missing $3^3P_1$
charmonium $\chi''_{c2}$ in the constituent quark model.  Besides, the
large mass difference does not support that the structures $Y(4274)$
and $Y(4320)$ are a same resonance.  Hence, the bump structure near
4.3 GeV reported by CMS collaboration and  CDF collaboration may be
composed of two structures. In fact, in Fig.~\ref{Fig: Exp} one can
find a hint of double-peak feature in $J/\psi\phi$ invariant mass
spectra by CMS and CDF collaborations (The peaks in CDF data are close
to the ones in CMS data with a translation about $10$~MeV). To verify
this assumption  we will fit the distribution of the mass difference, that is, the $J/\psi \phi$ invariant mass spectrum, released by
CMS collaboration recently with two resonances $Y(4274)$ and
$X(4320)$. It will be also helpful to understand  the absence of
$X(4350)$ in the $B$ decay process $B^+\to J/\psi \phi K^+$.

This work is organized as follows. In the next section we will study
the bound state of $D_sD_{s0}(2317)$ system through solving the BS
equation. The two body decay pattern of $Y(4274)$ will be calculated
with the wave function obtained. In the section 3 the invariant mass
spectrum is analyzed. In the last section, a summary and discussion
will be given.

\section{$Y(4274)$ as $D_sD_{s0}(2317)$ molecular state}

In this section we will study whether the $D_sD_{s0}(2317)$ system can
generate a bound state which decay pattern consistent with the
experimental observation of $Y(4274)$ by solving the BS equation.  Due
to the complication of direct solution of the BS equation, we will
apply two popular forms of 3-dimension reduction,
Blankenbecler-Sugar-Logunov-Tavkhelidz (BSLT) and Gross
formalisms~\cite{Nieuwenhuis:1996mc,Blankenbecler:1965gx,Logunov:1963yc,Gross:1982nz}
to find the solution of the BS equation. With the wave functions
obtained, the decay widths of $Y(4274)$ in $\gamma\gamma$ and $J/\psi
\phi$ channels can be calculated.

\subsection{The bound energy of $D_sD_{s0}(2317)$ system}

The 3-dimension BS equation for the normalized wave function
$|\phi\rangle$  can be written as
\begin{eqnarray}
(W-E_1-E_2)\phi=\int \frac{d^3k'}{(2\pi)^3}\mathfrak{F}V\phi,
\quad {\rm with}  \left\{
		\begin{array}{l}
\mathfrak{F}_{G}=1,\\
\mathfrak{F}_B=\frac{\sqrt{4(E_1+E_2)(E'_1+E'_2)}}{E_1+E_2+W},
		\end{array}
		\right.\ \ \ \ \ \ \ \label{Eq: 3DBS}
\end{eqnarray}
where $W$, $E_1=\sqrt{{\vec k}^2+m_1^2}$, $E_2=\sqrt{{\vec k}^2+m_2^2}$,  are the energies of $D_sD_{s0}(2317)$ system, $D_s$ and $D_{s0}(2317)$, respectively. The reduced potential $V=i{\cal V}/{\sqrt{2E_12E_22E'_12E'_2}}$
, which is reduced to the usual one-boson-exchange
model after non-relativization~\cite{Liu:2010hf,He:2011ed}.
The $\mathfrak{F}_G$ and $\mathfrak{F}_B$ are for Gross and BSLT 
formalisms, respectively. The explicit derivation can be found in
~\ref{Sec: 3D}.

The potential with light meson exchanges can be obtained with the Lagrangian from the heavy quark field theory~\cite{Casalbuoni:1996pg},
\begin{eqnarray}\label{eq:lag-p-exch}
 \mathcal{L}&=&
  i\frac{h}{f_\pi}(P^\dag_a\overleftrightarrow{\partial}^\mu P^*_{0b}
   + P^{*\dag}_{0a}\overleftrightarrow{\partial}^\mu P_b)\partial^\mu{}\mathbb{P}_{ba}
-i\frac{\beta{}g_V}{\sqrt{2}}
  P_a^\dag\overleftrightarrow{\partial}^\mu P_b\mathbb{V}^\mu_{ba}
  +  i\frac{\beta'g_V}{\sqrt{2}}P^{*\dag}_{0a}\overleftrightarrow{\partial}^\mu P^*_{0b}
  \mathbb{V}^\mu_{ba},\label{lagrangian}
\end{eqnarray}
where the coupling constants
$h=-0.56\pm0.28$,
$\beta\beta'=0.90$, $g_V=m_\rho/f_\pi=5.8$ with $f_\pi=132$
MeV~\cite{Liu:2010hf,He:2011ed,Casalbuoni:1996pg,Isola:2003fh}.
Since $\beta$ and $\beta'$ are not well determined in the literature.
Different values $\beta\beta'=0.9\eta_{\beta\beta'}$ with
$\eta_{\beta\beta'}=1,2$ will be
considered in this work.

The annihilation operations
$P,\,P^*_\mu,\,P_0^*,$ and $\,P_{1\mu}^\prime$
satisfy the normalization relations
\begin{eqnarray}
\langle 0|P|Q\bar{q}(0^-)\rangle=1,\
\langle
0|P^*_\mu|Q\bar{q}(0^-)\rangle=\epsilon_\mu,\
\langle 0|P_0^*|Q\bar{q}(0^+)\rangle=1,\ \langle
0|P_{1\mu}^\prime|Q\bar{q}(1^+)\rangle=\epsilon_\mu.
\end{eqnarray}
In Eq. (\ref{lagrangian}), the octet pseudoscalar and nonet vector
meson matrices read as
\begin{eqnarray}
\mathbb{P}&=&\left(\begin{array}{ccc}
\frac{\pi^{0}}{\sqrt{2}}+\frac{\eta}{\sqrt{6}}&\pi^{+}&K^{+}\\
\pi^{-}&-\frac{\pi^{0}}{\sqrt{2}}+\frac{\eta}{\sqrt{6}}&
K^{0}\\
K^- &\bar{K}^{0}&-\frac{2\eta}{\sqrt{6}}
\end{array}\right),\ \ \mathbb{V}=\left(\begin{array}{ccc}
\frac{\rho^{0}}{\sqrt{2}}+\frac{\omega}{\sqrt{2}}&\rho^{+}&K^{*+}\\
\rho^{-}&-\frac{\rho^{0}}{\sqrt{2}}+\frac{\omega}{\sqrt{2}}&
K^{*0}\\
K^{*-} &\bar{K}^{*0}&\phi
\end{array}\right).\label{vector}
\end{eqnarray}

\begin{figure}[h!]
\begin{center}
\includegraphics[bb=60 540 630 720,scale=0.7,clip]{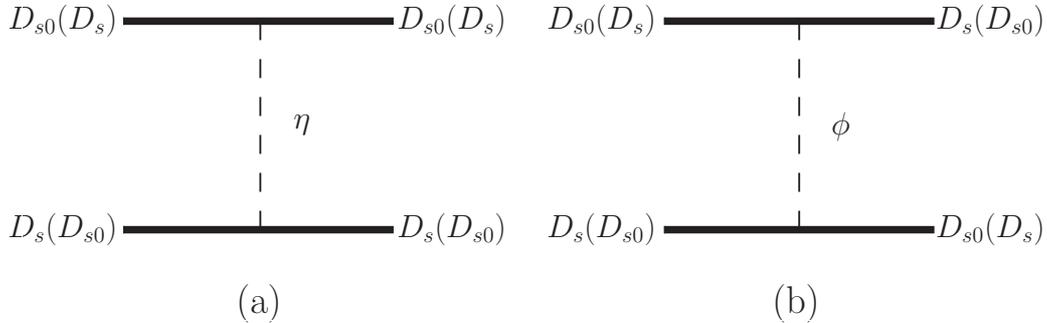}
\end{center}
\caption{The diagram for the interaction of  $D_sD_{s0}(2317)$ system by exchanging $\eta$ meson (a) and
$\phi$ meson (b).
	\label{Fig: potential}}
\end{figure}

The interaction mechanism for the $D_sD_{s0}(2317)$ is shown in
Fig.~\ref{Fig: potential}. Only pseuadoscalar light meson $\eta$ and
vector light meson $\phi$ exchange are possible because the vertices
of charm-strange meson and light meson with $u/d$ quark are suppressed
by OZI rule.

The explicit form of the kernel ${\cal V}={\cal V}_\phi+2/3{\cal V}_\eta$ can be written as
\begin{eqnarray}
	{\cal V}_\phi&=&-i\frac{\beta\beta'
	g^2_V}{2}~(p_i^\mu+p^\mu_f)
\frac{-g^{\mu\nu}+q^\mu q^\nu/m_\phi^2}{q^2-m_\phi^2}
(p'^\nu_i+p'^\nu_f),
\nonumber\\
{\cal
V}_\eta&=&-\frac{h^2}{f^2_\pi}~(p_i+p_f)\cdot q
\frac{i}{q^2-m_\eta^2}(p'_i+p'_f)\cdot q,\label{Eq: potential}
\end{eqnarray}
where $p_{i,f}$ and $p'_{i,f}$ are the momenta of the particle 1 and 2 in
the initial (final) state and $q=p_f-p_i$. $m_\phi$ and $m_\eta$ are
the masses of exchanged mesons $\phi$ and $\eta$.
A monopole form factor $F(q^2)=(\Lambda^2-m^2)/(\Lambda^2-q^2)$ is introduced to compensate the
off-shell effect of the changed light meson.

\begin{figure}[h!]
\begin{center}
\includegraphics[bb=80 20 950 780 ,scale=0.43, clip]{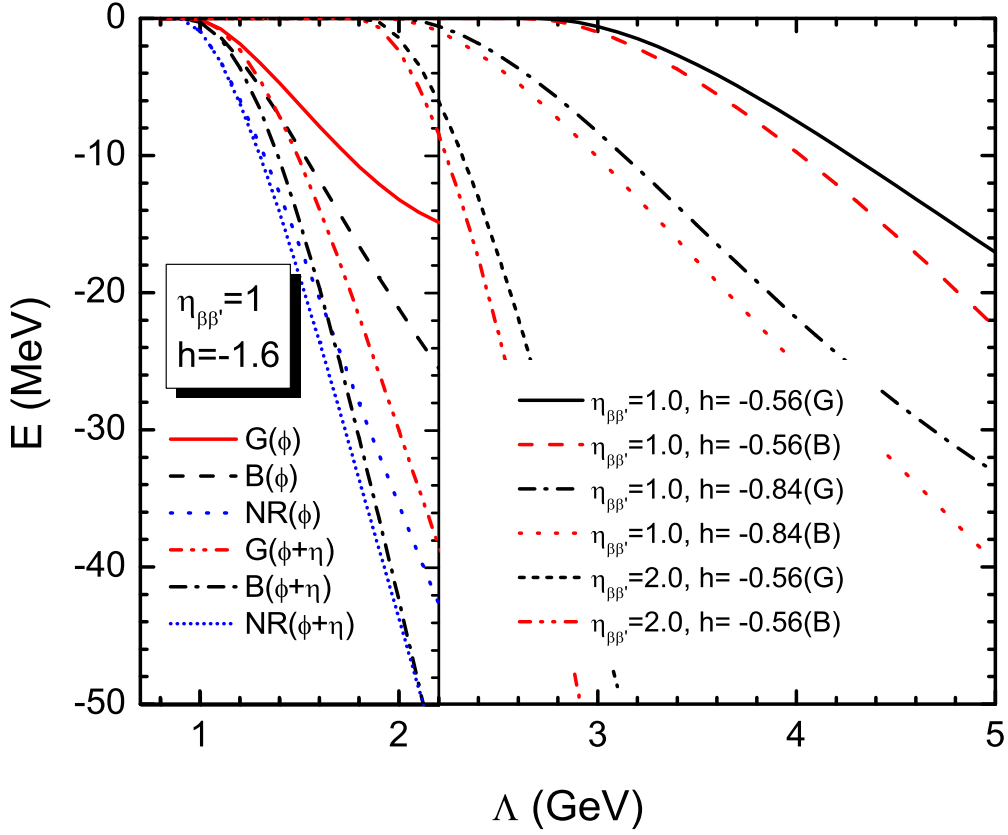}
\end{center}
\caption{The bound energy for $D_sD_{s0}$ system with different cutoff
	$\Lambda$ and coupling constants. Here
	$\beta\beta'=0.9~\eta_{\beta\beta'}$. NR, G and B stand for
	non-relativistic model, BS equation with Gross formalism and
	BS equation with BSLT formalism. \label{Fig: E}}
\end{figure}

With the potential (\ref{Eq: potential}), the 3-dimension BS equation
~(\ref{Eq: 3DBS}) can be solved numerically with the recursion
method. The explicit can be found in \ref{Sec: NS}. Since the
parameters $\beta\beta'$ and $h$ is not well determined in the
literature. We will present the results with different values of parameters.
The obtained bound energies  $E=W-m_{D_s}-m_{D_{s0}}$ are shown in Fig.~\ref{Fig: E}.

In the left part of Fig.~\ref{Fig: E}, the results by the
non-relativistic (NR) one-boson exchange potential in Refs.~\cite{Liu:2010hf,He:2011ed} are given. We find
the results obtained with the numerical method in this work is same to
the original model~\cite{Liu:2010hf,He:2011ed}, which can be seen as a
verification of our numerical method. In NR model the contribution
from $\eta$ exchange is negligible for reasonable cut-off $\Lambda$ (
the cut-off $\Lambda$ should be in the region 1$\sim$5 GeV). So only
the parameter $h$ is adjusted to find a cut off $\Lambda$, with which
the bound energy is 13 MeV as suggested by experiment, and a value
$h=-1.6$ is adopted in Refs.~\cite{Liu:2010hf,He:2011ed}.

However, the $h$ is better determined with value $h=0.56\pm0.28$ obtained by
the sum rule \cite{Casalbuoni:1996pg}  than $\beta\beta'$.   Moreover,
as shown in the left part
of Fig.~\ref{Fig: E} the contributions from $\eta$ exchange become
more important in BS approach, which is relativistic, than in NR model.
With a non-zero contribution from $\eta$ exchange, the solution
can be found in BS approach if we adopt $h=0.56\pm0.28$ obtained by
the sum rule\cite{Casalbuoni:1996pg} as shown in the right part of
Fig.~\ref{Fig: E}. In the NR model, no solution with bound energy
about 10 MeV can be found with such value of $h$. If adopting a larger
$\beta\beta'=1.8$ the solution with bound energy about 10~MeV can be
found with a cut-off $\Lambda$ about $2.5$~GeV. As shown in
Fig.~\ref{Fig: E} with all parameters considered here, a loose bound
state can be found when solving BS equation. Hence, a molecular state $Y(4274)$ can be generated from
$D_sD_{s0}(2317)$ system.

From Fig.~\ref{Fig: E}, the results from Gross formalism, BSLT
formalism and NR model are close to each other for the small bound
energy. The solution with bound energy near zero appears with the
almost same cut off $\Lambda$. In other word, if we focus on the
molecular state with very small bound energy, the three models give a
very similar results. If a deep bound molecular state is considered,
more complicated formalism should be adopted.

\subsection{The decay pattern of $Y(4274)$}

\begin{figure}[h!]
\begin{center}
\includegraphics[bb=80 370 540 725,scale=0.43,clip]{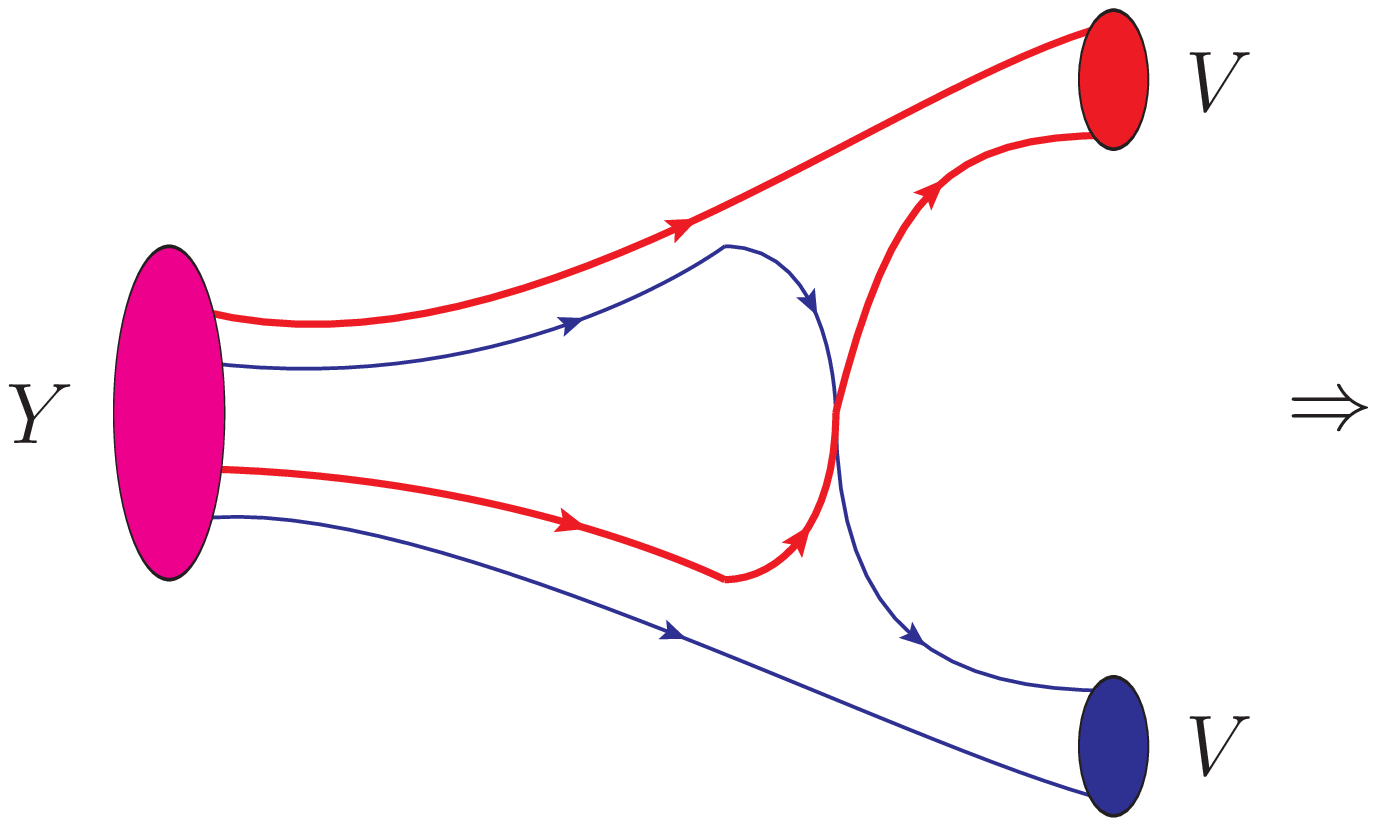}
\includegraphics[bb=160 440 440 725,scale=0.7,clip]{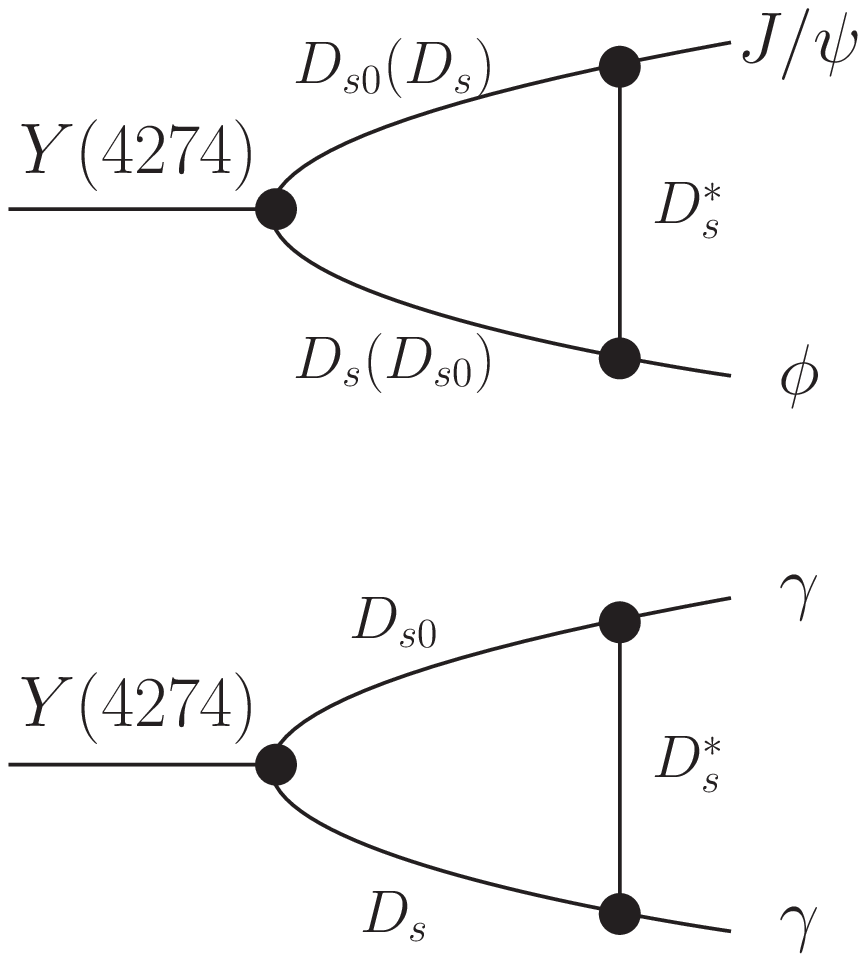}
\end{center}
\caption{Decay mechanism of $Y(4274)$ to $J/\psi\phi$ and
	$\gamma\gamma$. The left diagram is in the quark level and right diagrams are in hadron level.  \label{Fig: mechanism}}
\end{figure}

Now we have obtained the wave function, which
contain the information of vertex, as well as the bound energy.  We will estimate the decay widths of
$Y(4274)$ to $J/\psi \phi$ and $\gamma\gamma$ with the wave function.
Assuming $Y(4274)$ is a molecular state, the dominant decay mechanism
is exchanging  charm/light quarks in the two constituent mesons as shown in
Fig.~\ref{Fig: mechanism}.

The decay mechanism in quark level can be described by a hadronic
loop, and the amplitudes can be written as
\begin{eqnarray}
	{\cal M}&=&\Gamma G A =\Gamma (g+\Delta G)A \approx \Gamma g A	\equiv\int \frac{d^3 k}{(2\pi)^4}\psi(|{\vec k}|)A(|{\vec
	k}|),
\end{eqnarray}
where the $\Gamma$, $G$ and $A$ are the vertex for the $Y(4274)$
decaying to $D_s$ and $D_{s0}(2317)$, propagators of $D_s$ and $D_{s0}(2317)$,
and the amplitudes for  $D_s$ and $D_{s0}(2317)$ to $J/\psi\phi$ and
$\gamma\gamma$ by exchanging $D^*_s$. Here the term with $\Delta G$ is omitted as usual.

\begin{figure}[h!]
\begin{center}
\includegraphics[bb=30 20 760 590 ,scale=0.55, clip]{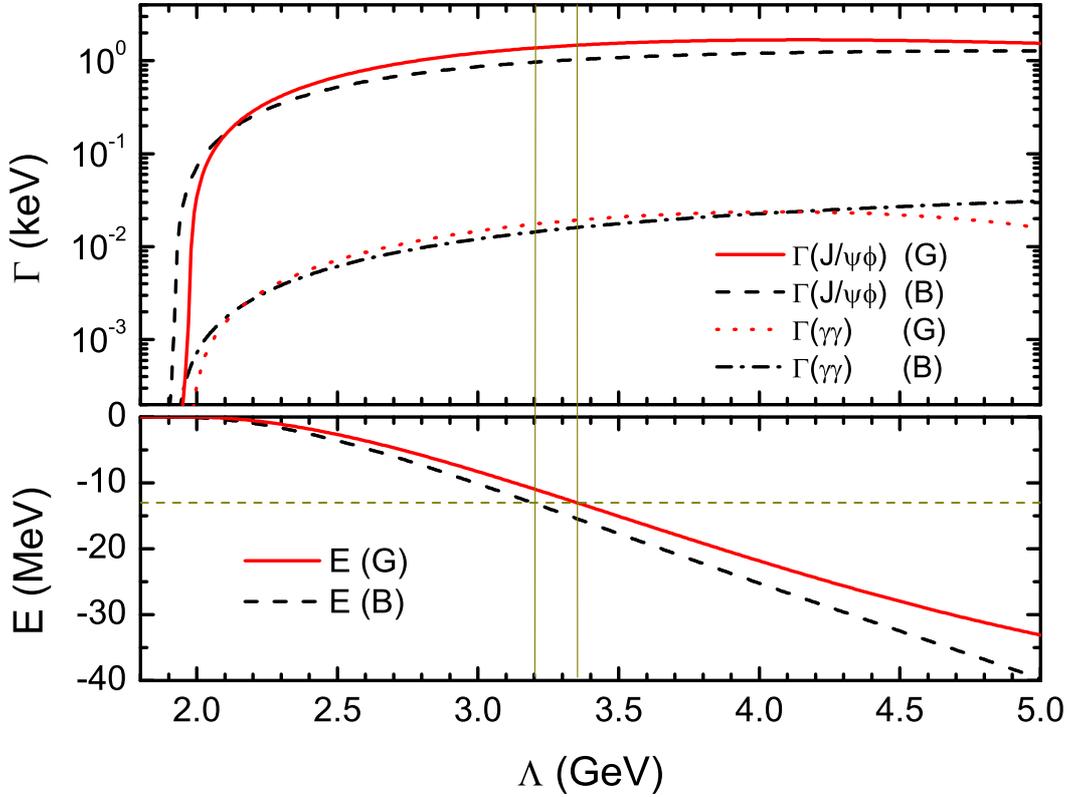}
\end{center}
\caption{The decay widths for $Y(4274) \to J/\psi\phi/\gamma\gamma$  and bound energy with $h=0.84$ and
	$\eta_{\eta\eta'}=1$. The upper subfigure is for the decay widths of $Y(4274)$. The lower subfigure is the bound energy obtained with the same parameter. \label{Fig: width}}
\end{figure}

The Lagrangian  used for $A_{J/\psi\phi}$ are~\cite{Casalbuoni:1996pg,Colangelo:2003sa},
\begin{eqnarray}
	{\cal L}&=&-2g_3\sqrt{m_{D}m_{D^*}m_\psi}
\psi\cdot P^*_a~P_{0a}-i2 g_2 \sqrt{m_\psi m_{D_{(s)}}
	/m_{D^*_{(s)}}}\nonumber\\
&\cdot&\varepsilon_{\beta \mu \alpha \tau}
\partial^\beta \psi^\mu (P^\dag_a\overleftrightarrow{\partial}^\tau P_a^{*\alpha}
+P^{*\alpha\dag}_a\overleftrightarrow{\partial}^\tau P_a)\nonumber \\
 &-&i\sqrt{2}\lambda{}g_V\varepsilon_{\lambda\alpha\beta\mu}
 (P_a^{*\mu\dag}\overleftrightarrow{\partial}^\lambda
 P_b+P_a^{\dag}\overleftrightarrow{\partial}^\lambda P_b^{*\mu})
 \partial^\alpha \mathbb{V}^\beta_{ab}\nonumber\\
 &-&\sqrt{2}\zeta{}g_V\sqrt{m_{D_0}m_{D^*}}(P^{*\dag}_{0a} P_b^{*\mu}+
  P^{*\mu\dag}_{a}P^*_{0b})\mathbb{V}_{\mu{}ba}\nonumber\\
  &\pm&\sqrt{2}\varpi g_V(P^{*\dag}_{0a}\overleftrightarrow{\partial}^\alpha P^{*\beta}_b -
  P^{*\beta\dag}_{a}\overleftrightarrow{\partial}^\alpha
  P^*_{0b})(\partial_\alpha{}\mathbb{V}_\beta -
  \partial_\beta{}\mathbb{V}_\alpha)_{ba},\ \ \ \ \ \ \ \
\end{eqnarray}
where $g_3={\sqrt {m_\psi}/f_\psi}$ and
$g_2=\sqrt{m_\psi}/(2m_Df_\psi)$ with $f_{J/\psi}=405\pm14$~MeV, $\lambda=0.56$ ~GeV$^{-1}$, $\zeta=0.727$ and $\varpi=0.364$
~\cite{Casalbuoni:1996pg,Colangelo:2003sa}.
For the two photon decay,
\begin{eqnarray}
	{\cal L}&=&g_{D_{s0}D_{s0}\gamma}D_{s0}A^{\mu\nu}F^{\mu\nu}
	+g_{D^*_{s}D_s\gamma}D_s\epsilon_{\mu\nu\alpha\beta}
	A^{\mu\nu}  F^{\alpha\beta}.
\end{eqnarray}
The coupling constant can be obtained by the decay widths. The radiative
decay of $D_{s0}(2317)$ to $D^*_s$ have been calculated theoretically
in literature,  here we adopt a typical value
$\Gamma^\pm_{D_{s0}D^{*\pm}_s\gamma}=1$~keV~\cite{Close:2005se}. For the decay of $D^*_s$
to $D_s$, we choose the upper limit of the values in PDG as $1.9\times94.2\%$~MeV~\cite{PDG}.
The decay widths with the variation of the cut-off $\Lambda$ are shown in
Fig.~\ref{Fig: width}.

An interesting phenomena can be found that the decay width increase
with the increases of bound energy $|E|$, which is different from the
case of three body decay~\cite{He:2011ed,He:2012zd} where the decay
width decreases.  For a molecular state, which is a bound state of consistent hadrons, a large bound energy (which also means small radius of the molecular state as many calculation suggested, for example Ref. \cite{He:2012zd}) will be helpful to the occurrence  of the exchanging of quarks in the constituents, which is the main mechanism of the two body decay.  For the three body decay, the main decay mechanism is the collapse of the molecular state in sequential
three-body decay, a looser bound system is beneficial to the collapse of the constituents. Hence our results in the two and three body decays of $Y(4274)$ reflect the internal structure of the molecular state.

\section{The $J/\psi \phi$ invariant mass spectrum}

As shown in the previous section, a bound state  can be
generated from the $D_sD_{s0}(2317)$ system. The theoretical results
about $J/\psi \phi$ and $\gamma\gamma$ decays of this state are consistent with the
experimental observed structure $Y(4274)$. However, In the CMS
experiment~\cite{ATLAS:2013eta}, $Y(4274)$ was not reported while a
structure $X(4320)$, which is about 50 MeV higher than $Y(4274)$, was
reported.  In this
section, we will analysis the invariant mass spectrum released by CMS
collaboration to answer the questions proposed in introduction.

\begin{figure}[h!]
\begin{center}
\includegraphics[bb=130 560 470 720,scale=0.9,clip]{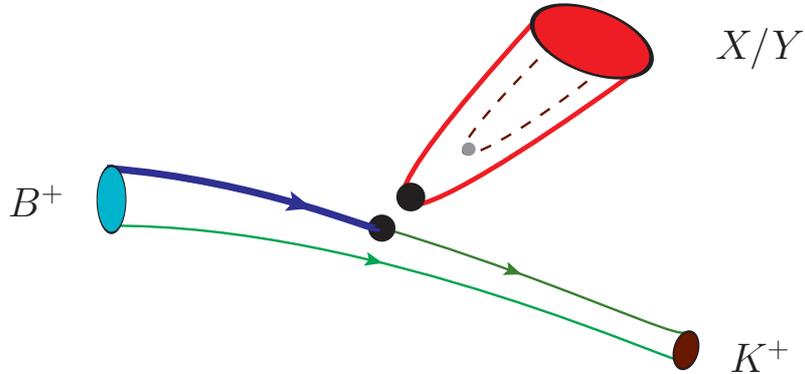}
\end{center}
\caption{Weak decay mechanism  of $B^+\to K^+
	X/Y$.\label{Fig: weakdecay}}
\end{figure}

Now we make an analysis about the decays of $B^+$ to $X/Y$s.  The
leading order of the weak decay can be described as a four-quark
local interaction by the effective
Hamiltonian~\cite{Colangelo:2003sa,Zanetti:2011ju} as shown in
Fig.~\ref{Fig: weakdecay},
\begin{eqnarray}
	{\cal
	H}=\frac{G_F}{\sqrt{2}}V_{cb}V^*_{cs}\left[(C_2+\frac{C_3}{3}){\cal
	O}_2+\cdots\right],\label{Eq: Hweak}
\end{eqnarray}
where ${\cal O}_2=(\bar{c}\Gamma_\mu c)(\bar{s}\Gamma^\mu b)$ with
$\Gamma_\mu=\gamma_\mu(1-\gamma_5)$.

The decay amplitude can be factorized by splitting the matrix into two
pieces,
\begin{eqnarray}
	{\cal A}&=&iF\langle  B(p)|J^W_\mu|K(p')\rangle\langle X(q)|J^{\mu
	(cc)}|0\rangle,
\end{eqnarray}
with $F=\frac{G_F}{\sqrt{2}}V_{cb}V_{cs}^*(C_2+\frac{C_1}{3})$.
The $B$ to $K$ part can be described as,
\begin{eqnarray}
\langle
B(p)|J^W_\mu|K(p')\rangle=f_+(q^2)P_\mu+f_-(q^2)q_\mu.
\end{eqnarray}
The explicit form and the parameterizations  for $f_\pm(q^2)$ can be
found in Ref.~\cite{Melikhov:2000yu}.

The matrix $\langle X|J|0\rangle$  for pseudoscalar and axial vector
meson can be parameterized with decay
constant as
\begin{eqnarray}
\langle X_{0^{-+}}(q)|J^{\mu (cc)}|0\rangle&=&-if_P q^\mu,\\
\langle X_{1^{++}}(q)|J^{\mu
(cc)}|0\rangle&=&f_Am_A\epsilon_\mu^*(q),
\end{eqnarray}
where $f_A$, $f_P$, $m_A$ and $\epsilon_\mu$ are the decay constants for pseudoscalar and axial vector meson, mass of axial vector meson, and polarized vector of the axial vector meson, and $q$ are the momentum of the meson.

The $X(4350)$ observed in two photon fusion have been suggested as
a $3^3P_2$ charmonium state $\chi''_{c2}$~\cite{Liu:2009fe}, which vanishes in the $B$ decay in the
factorization approximation. Such suppression is obvious in the
production of $1^3P_J$ states $\chi_{cJ}$ in $B\to \chi_{cJ} K^+$ channel~\cite{PDG}.
The observation of $X(4320)$ in $B$ decay suggested it should not be
$X(4350)$ found in the two photon fusion.  The P-wave charmonium
$\chi''_{c1}$ should have a mass close to $\chi''_{c2}$, and $X(4320)$
observed by CMS collaboration have a mass close to the value $4317$~MeV
predicted in constituent quark model ~\cite{Barnes:2005pb} as shown in
Fig~\ref{Fig: Exp}. Moreover as an axial vector meson it can not be
produced in the two photon fusion as the Belle experiment suggested
while the production of $\chi''_{c1}$ should be considerable in the
$B$ decay. Hence it is reasonable to assign $X(4320)$ as the missing
$\chi''_{c1}$.  The absence of $Y(4140)$ in the two photon fusion
indicts it should not be a molecular
state~\cite{Shen:2009vs,Branz:2009yt}, which is also disfavored
by the large bound energy. Here, we adopt the assignment of $Y(4140)$ as an tetraquark
$c\bar{c}s\bar{s}$ with $J^P=1^{++}$, which can explain the absence in two
photon fusion and its mass and decay pattern~\cite{Stancu:2009ka}.

Hence, in the $J/\psi\phi$ invariant mass spectrum from $B$ decay there
exist three resonances,
$Y(4140)$ with $1^{++}$, $X(4320)$ with $1^{++}$ and $Y(4274)$ with
$0^{-+}$. The Lagrangians for the sequential decays of the pseudoscalar and axial
vector mesons $X/Y$ to
the $J/\psi\phi$ can be written as
\begin{eqnarray}
	{\cal
	L}&=&\frac{g_P}{M_X}i\epsilon_{\mu\nu\alpha\beta}\psi^{\mu\nu}\phi_{\alpha\beta} X
	+g_Ai\epsilon_{\mu\nu\alpha\beta}\partial^\mu X^\nu
	\tilde{\psi}^\alpha\tilde{\phi}^{\beta}, \label{Eq: XYdecay}
\end{eqnarray}
where $\phi^{\alpha\beta}=\partial^\alpha\phi^\beta-\partial^\beta
\phi^\alpha$, and $\tilde{\phi}^\beta=(g^{\beta\rho}-k_\phi^\beta k_\phi^\rho
/m_\phi^2) \phi_\rho$ to keep the gauge invariance.  $\psi^{\mu\nu}$ and $\tilde{\psi}^\alpha$ for
$J/\psi$ is analogous. $g_P$ and $g_A$ are the coupling constants for  pseudoscalar and axial
vector mesons, respectively.

The amplitude ${\cal M}_{X/Y}$ for $X/Y$ can be obtained from
Eqs.~(\ref{Eq: Hweak}-\ref{Eq: XYdecay}) and the propagator of $X/Y$ which
involves Breit-Wigner mass $m$ and width $\Gamma$. With a constant background ${\cal L}_{bk}=Cg^{\mu\nu}$, the square of the amplitude can be
written as,
\begin{eqnarray}
	|{\cal M}|^2=4C^2+|{\cal M}_{Y(4140)}|^2+|{\cal
	M}_{Y(4274)}|^2+|{\cal M}_{X(4320)}|^2
	+2{\rm Re}(e^{i\phi}{\cal
	M}^{\mu\nu}_{Y(4140)}{\cal
	M}^*_{X(4320)\mu\nu}).\ \  \label{Eq: amplitudes}
\end{eqnarray}
One can find only one interference term are left.

With the square of the amplitude, the $J/\psi \phi$ invariant mass spectrum for
process $B^+\to K^+ J/\psi \phi$ can be calculation. The parameters
are the strength constants $N$, Breit-Wigner masses $m$ and widths
$\Gamma$ of $Y(4140)$, $Y(4274)$ and $X(4320)$. Here strength constant
$N$ involve the coupling constants of corresponding resonance and a general
normalization.  Now, we determine the parameters by fitting the
distribution of mass difference $m(\mu^+\mu^--K^+K^-)-m(\mu^+\mu^-)$, this is, the
$J/\psi \phi$
invariant mass spectrum, released by CMS collaboration by
binned maximum likelihood~\cite{PDG} with three resonances. It is done
through minimizing $-2\ln{\cal
L}=2\sum_i^N[\nu_i-n_i+n_i\ln\frac{n_i}{\nu_i}]$
where $n_i$ and $\nu_i$ are the experimental and theoretical values in
$i^{\rm th}$ bin using the
MINIUT code. The  results are shown in Fig.~\ref{Fig:
fit}.
\begin{figure}[h!]
\begin{center}
\includegraphics[bb=71 11 390 319 ,scale=1.1, clip]{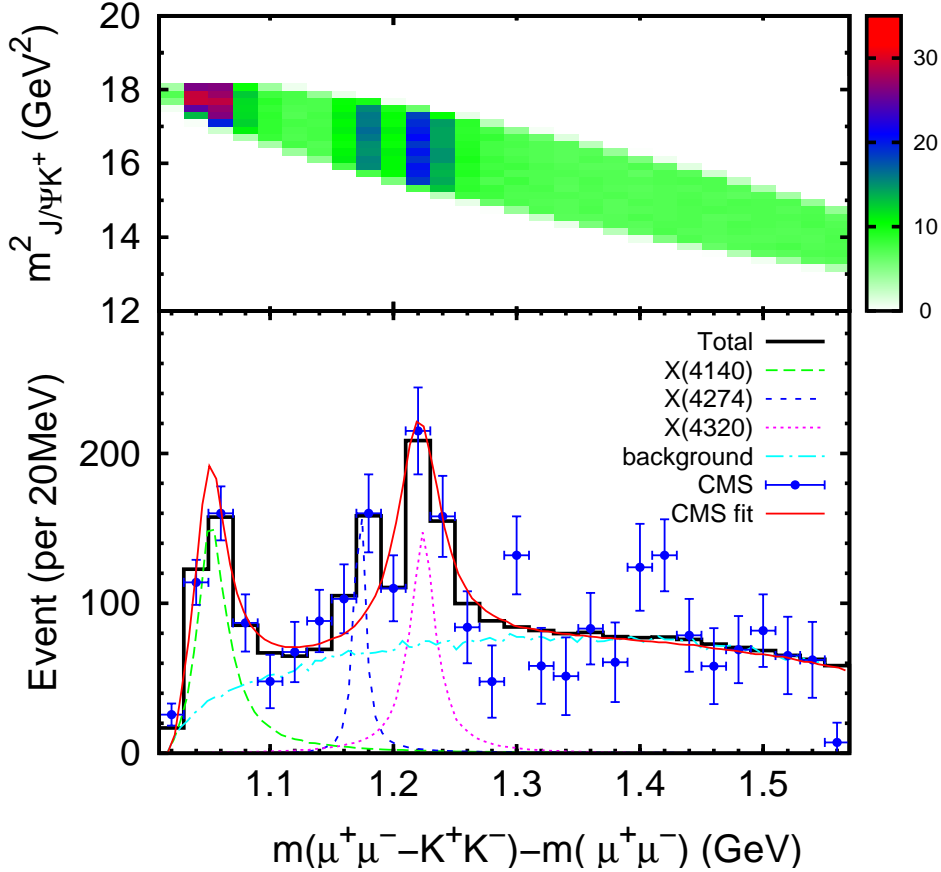}
\end{center}
\caption{Distribution of the mass difference $m(\mu^+\mu^--K^+K^-)-m(\mu^+\mu^-)$ and
	Dalitz plot for $B^+\to J/\psi\phi K^+$. The histogram
	(black) is for
	the fitting results in the current work. The long dashed
	(green), short dashed( blue), dotted (magenta) and dash dotted
	(cyan) lines are for the contributions from $X(4140)$,
	$X(4274)$, $X(4320)$ and background. The full circle (blue)
	and full line (red) are for the experimental data
	and the fitting results of CMS
	Collaboration~\cite{ATLAS:2013eta}.
	\label{Fig: fit}}
\end{figure}

Different from the two Breit-Wigners fitting in the experimental
article~\cite{ATLAS:2013eta}, here we use three Breit-Wigners. One can
found the CMS data is well fitted with three Breit-Wigners especially
below $m(\mu^+\mu^--K^+K^-)-m(\mu^+\mu^-)=$1.3~GeV. The first structure around
1.05 GeV is about 0.15 GeV below
the second structure around 1.2 GeV, and dominant with the
contribution of resonance $Y(4140)$. Hence, the fitting result in the
current work is close to that in Ref.~\cite{ATLAS:2013eta}.  For the
second structure, only one Breit-Wigner are used in the fitting of the
CMS collaboration, and a resonance with $m=4316.7$ MeV was reported. In
the current work, this structure is fitted with two resonances,
$Y(4274)$ and $X(4320)$, which results in an excellent reproduction of
the experimental distribution of the mass difference around 1.2 GeV.
The Dalitz plot for $B^+\to J/\psi\phi K^+$ is also presented for
reference, which is obtained by FOWL program in CERNLIB.

The explicit about the fitted parameters is shown in Table~\ref{Tab: para}.
The Berit-Wigner masses of $Y(4140)$, $Y(4274)$ and $X(4320)$ obtained in the current
fitting are 4147.8 MeV, 4269.4 MeV and 4320.8 MeV, which are close to the values
suggested in the original experimental
articles as shown in Fig.~\ref{Fig: Exp}\cite{ATLAS:2013eta,Yi:2010aa,Shen:2009vs}. The Breit-Wigner
width of $Y(4274)$ is 9.4 MeV, which is much smaller than the value
of PDF collaboration, 32.3$^{+21.9}_{-15.3}$(stat)$\pm$7.6(syst) MeV.
It is easy to understand because two resonances $Y(4274)$ and
$X(4320)$ are considered in the current work. The total decay width of $X(4320)$ is consistent to
the constituent quark model and $^3P_0$ model predictions with a assignment of
$\chi''_{c1}$~\cite{Barnes:2005pb,Liu:2009fe}.

\renewcommand\tabcolsep{0.8cm}
\renewcommand{\arraystretch}{1.5}

\begin{table}[h!]
\begin{center}
\begin{tabular}{c|rrrrrr}
\hline
& {$Y(4140)$} & $Y(4274)$ &$X(4320)$ \\
\hline
$N$ & 0.139$\pm0.007$  & 0.069$\pm0.006$ &0.094$\pm0.007$ \\
$m$ & 4147.8$\pm 2.6$  & 4269.4$\pm1.5$ &4320.8$\pm1.6$ \\
$\Gamma$ & 29.6$\pm 2.3$  & 9.4$\pm 1.4$ & 25.0$\pm 1.7$
\\ \hline
Sig & 13.3[11.7] $\sigma$  & 4.7[5.2] $\sigma$  &$6.2[8.4]$ $\sigma$
\\ \hline
\end{tabular}
\end{center}
\caption{The fitted strength constant $N$, Breit-Wigner masses $m$
	and widths $\Gamma$ for the three resonances. The best
	fitted phase angle $\phi=0.23\pm0.09$ and the background
	constant $C=1.98\pm0.04$. The last line is the significance
	$-2\Delta \ln {\cal L}$ by taking off the corresponding
	resonance in the full model or adding a resonance into
	background (with bracket). \label{Tab: para}
}
\end{table}

To present the importance of each resonance in fitting, we also give
the significance $-2\ln({\cal L'}/{\cal L})$ of the binned maximum
likelihood. Here ${\cal L}$ and ${\cal L}'$ are for the full
model and changed model, respectively.  By turning off the
corresponding resonance from the full model, we find the significances
of $Y(4274)$ and $X(4320)$ are 4.7~$\sigma$ and 6.2~$\sigma$ by
turning of the  the corresponding resonance in the full model or
5.2~$\sigma$ and 8.4~$\sigma$ by adding a resonance into background.
Hence both $Y(4274)$ and $X(4320)$ are important to fit the
distribution of mass difference.

\section{Summary and discussion}

In this work, we study the mass and the decay pattern of the $Y(4274)$ as a molecular state $D_sD_{s0}(2317)$ in the BS equation approach with quasipotential. The solution with reasonable parameters is found in the $D_sD_{s0}(2317)$ interaction by exchanging the light meson $\eta$ and $\phi$. The absence of the $Y(4274)$ in the $\gamma\gamma\to J/\psi \phi$ process can be explained by the decay widths of $Y(4272)$ in $\gamma\gamma$ and $J/\psi\phi$ channels, which are calculated with the wave function obtained in the solution of the BS equation.

The assignment of $X(4320)$ observed in $B$ decay and $X(4350)$
observed in two photon fusion as $P$-wave charmoniums $\chi''_1~ (3^3P_1)$  and $\chi''_2~(3^3P_2)$ is consistent to
existing experiment observations and the theoretical
predictions about mass and decay width.
Since the factorization approximation suggests suppression of
$\chi''_{c0}$ compared with $\chi''_{c1}$ in $B^+\to
J/\psi \phi K^+$, the assignment of $X(4320)$ as $\chi''_1$ excludes the possibility to assign $Y(4274)$ as
$\chi''_{c0}$ combined with its considerable contributions in $B$ decay. It is also supported by the small width of $Y(4274)$, which conflict with the prediction in the $^3P_0$ model~\cite{Liu:2009fe}.

Based on the calculation and analysis in this work we can reach following conclusions.
\begin{itemize}
  \item The bump structure near 4.3 GeV in the $J/\psi\phi$ invariant mass spectrum in the $B^+\to
J/\psi \phi K^+$ is from two resonances, $Y(4274)$ as a  $D_sD_{s0}(2317)$ molecular state and $X(4320)$ which can be assigned as $\chi''_{c1}$.
  \item The absence of $Y(4274)$ in the two photon fusion can be explained by the decay pattern of $Y(4274)$.
  \item The structure $X(4350)$ found in two photon fusion can be assigned as $\chi''_{c2}$, which contribution should be suppressed in $B^+\to
J/\psi \phi K^+$.
\end{itemize}

The more precise
data from forthcoming BelleII and SuperB combined with explicit
theoretical studies of $X/Y$ production in $B$ decay may provide
clearer picture of double-peak feature and the internal structure
of these resonances.

\section*{Acknowledgment}

This project is partially supported by the National Natural Science
Foundation of China (Grants No. 11275235, No. 11035006, No. 10905077)
and the Chinese Academy of Sciences (the Special Foundation of
President under Grant No. YZ080425 and the Knowledge Innovation
Project under Grant No. KJCX2-EW-N01).

\appendix

\section{The 3-dimension reduction of BS equation}
\label{Sec: 3D}

The BS equation for the amplitudes ${\cal T}$ can be written as \cite{Adam:1997cx,Adam:1997rb,Garcon:2001sz}
\begin{eqnarray}
{\cal T}={\cal V}+{\cal V}G{\cal T},
\end{eqnarray}
where the propagator for the two particles $G=G_1~G_2=$ $i/(_1^2-m_1^2)~$ $i/(k_2^2-m_2^2)$ and ${\cal V}$ is the interaction kernel. For a bound state the amplitudes can be written as \cite{Adam:1997cx,Adam:1997rb,Garcon:2001sz}
\begin{eqnarray}
 {\cal T}=\frac{|\Gamma\rangle i\langle \Gamma|}{P^2-M^2},
\end{eqnarray}
where $|\Gamma\rangle$ is the vertex of bound state and two constituent particles, $P$ and $M$ are the momentum and mass of the bound state.

\begin{figure}[h!]
\begin{center}
\includegraphics[bb=0 600 770 750,scale=0.6,clip]{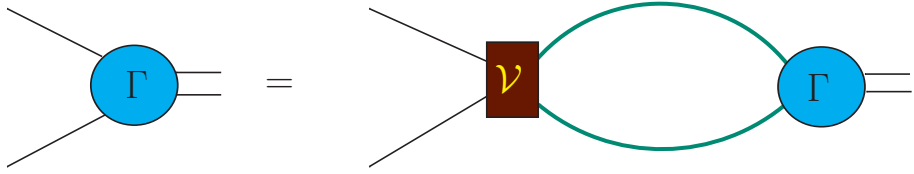}
\end{center}
\caption{The BS equation for the vertex $|\Gamma\rangle$. \label{fig: vertex}}
\label{BS_tr1}
\end{figure}

Hence, the BS equation for the vertex can be written as
\begin{eqnarray}
|\Gamma\rangle={\cal V}G|\Gamma\rangle,
\end{eqnarray}
with the normalization relation $1=\langle \Gamma|\partial G/\partial
P^2 |\Gamma|\rangle$ \cite{Adam:1997cx,Adam:1997rb,Garcon:2001sz}. It is figured in Fig.~\ref{fig: vertex}.

The vertex function $|\Gamma\rangle$ can be related to the BS
bound state wave function $|\psi\rangle$ as  \cite{Adam:1997cx,Adam:1997rb},
\begin{eqnarray}
	|\psi\rangle=G|\Gamma\rangle.
\end{eqnarray}

The BS equation for the vertex $|\Gamma\rangle$ can be rewritten as following form
\begin{eqnarray}
|\Gamma\rangle={\cal U}g|\Gamma\rangle
	, \ {\cal U}={\cal V}+{\cal V}\Delta G{\cal U},
\end{eqnarray}
 Here $\Delta G=G-g$, $g$ and $\cal U$ are
called the quasipotential two-body propagator and the quasipotential
kernel.
 As usual, we assume term with $\Delta G$ is small and can be neglected~\cite{Nieuwenhuis:1996mc,Blankenbecler:1965gx,Gross:1982nz,Adam:1997cx,Adam:1997rb}.

To satisfy the unitary condition, the propagator $g$ should satisfy the
relation
\begin{eqnarray}
g-g^\dag=2\pi i\delta((\eta_1(s)
P+k)^2-m^2)\delta((\eta_2(s) P-k)^2-m^2),\ \ \
\end{eqnarray}
where $\eta_1(s)+\eta_2(s)=1$ with $s=P^2$.
With $\epsilon_{1,2}(s)=(s+m_{1,2}^2-m_{2,1}^2)/2\sqrt{s}$, we can define
$\eta_{1,2}=\epsilon_{1,2}/(\epsilon_1+\epsilon_2)$.
Now we have many choice to write the propagator.
The most popular form is~\cite{Hung:2001pz,Zakout:1996md}
\begin{eqnarray}
	g&=&2\pi\int \frac{ds'}{s'-s+i\epsilon} h(s',s)~
	\delta([\eta'_1(s')P'+k]^2-m_1^2)~\delta([\eta'_2(s')P'-k]^2-m_2^2)
\end{eqnarray}
with $P'=\sqrt{s'/s}P$.

The choice of $h(s',s)$ is random to some extent. Here we adopt two widely used formalisms, BSLT and Gross formalisms~\cite{Nieuwenhuis:1996mc,Blankenbecler:1965gx,Gross:1982nz}. For BSLT formalisms, we choose $h(s'-s)=1$ and $\eta'(s')=\eta(s')$.
For Gorss formalism, $h(s',s)=(\sqrt{s'}+\sqrt{s})/\sqrt{s'}$ and
$\eta'_1(s')=\eta_1(s)\sqrt{s/s'}$ and
$\eta'_2(s')=1-\eta_1(s)\sqrt{s/s'}$.

The quasipotential propagators in the Gross (G) and BSLT (B) formalisms written down in the center of mass
frame where $P=(W,{\vec 0})$ are \cite{Nieuwenhuis:1996mc,Ramalho:2001pd}
\begin{eqnarray}
	g=\frac{-2\pi i}{2E_1E_2}
	\frac{f}{E_1+E_2-W},\quad {\rm with}  \left\{
		\begin{array}{l}
f_{G}=\delta(\epsilon_1+k^0-E_1),\\
f_{B}=\frac{(E_1+E_2)~\delta(k^0)}{E_1+E_2+W}.
		\end{array}
		\right.
\end{eqnarray}
It is easy to find that in Gross formalism the particle 1 is set on-shell and in BSLT formalism  $k^0=0$.
Due to the existence of the delta function, the BS equation in 4-dimension will be reduced into a 3-dimension equation.

The normalization of the vertex with quasipotential approximation in the center of mass frame has the following form,
\begin{eqnarray}
	1=\int \frac{d^4p}{(2\pi)^4}\Gamma^\dag\frac{\partial ig}{\partial W^2} \Gamma.
\end{eqnarray}
The normalized wave functions of bound state can be introduced
as $|\phi\rangle=N|\psi\rangle$ with
\begin{eqnarray}
	N_B=\frac{\sqrt{E_1E_2}}{\sqrt{(2\pi)^52(E_1+E_2)}},
        \ \
	N_G=\frac{\sqrt{2E_12E_2}}{\sqrt{(2\pi)^52W}},
\end{eqnarray}
and the wave function in 3-dimension normalized to $\int d^3p |\phi|^2=1$.

\section{Numerical solution of the  3-dimension BS equation}
\label{Sec: NS}

The 3-dimension BS equation Eq. \ref{Eq: 3DBS}  can be rewritten as
\begin{eqnarray}
W\phi({\vec k})&=&\int \frac{d^3{\vec
k}'}{(2\pi)^3}[\mathfrak{F}(W, {\vec k},{\vec k}')V(W, {\vec k},{\vec
k}')+(E_1({\vec k}')+E_2({\vec k}'))\delta({\vec k}'-{\vec k})]\phi({\vec
k}')\nonumber\\
&\equiv&\int \frac{d^3{\vec
k}'}{(2\pi)^3}~K(W, {\vec k},{\vec k}')\phi({\vec
k}'),\label{Eq: 3DBSre}
\end{eqnarray}
Here the wave function $\psi(\vec k)$ is radially symmetric and the
angular part are integrated out as refs. \cite{Guo:2007mm,Ke:2012gm}.
After defining the kernel ${\cal K}(W, {\vec k},{\vec k}')$ after integration as $A(W,
|{\vec k}|, |{\vec k}'|)=\int \frac{d\Omega'}{(2\pi)^3}~K(W, {\vec k},{\vec k}')$.
We reach a integral equation with form
\begin{eqnarray}
	W \psi(|{\vec k}|)=\int d~|{\vec k}'| A(W, |{\vec k}|, |{\vec
	k}'|) \psi (|{\vec k}'|.
\end{eqnarray}

To solve the integral equation, we discrete the $|{\vec k}|$ and $|{\vec
k}'|$ to $|{\vec k}|_i$ and $|{\vec k}|_j$ by the Gauss quadrature , then the
above equation transfer to a matrix equation
\begin{eqnarray}
	W\psi_i=\sum_jA_{ij}(W)\omega_j\psi_j\equiv
	\sum_j\tilde{A}_{ij}(W)\psi_j,
\end{eqnarray}
which can be written as a compact form
\begin{eqnarray}
	W\psi=\tilde{A}(W)\psi.
\end{eqnarray}

The integral equation involving a nonlinear dependence on the total
energy $W$ of the system reduce to a nonlinear spectral problem.  Here
we adopt the recursion method in
\cite{Soloveva2000,Soloveva2001,Skachkov2003}.  It proceeds by forming
a sequence of approximations to $W$ and $\psi$ using the recursion
relation:
\begin{eqnarray}
	W^{(l)}_n\psi=\tilde{A}(W^{(l-1)}_s)\psi,n=1,2,\cdots s,\cdots
\end{eqnarray}
where the upper index is the iteration number, and the
lower index is the eigenvalue number. At the first iteration step, an
input approximation of the sought eigenvalue is substituted into the
kernel.  In the problem here, we choose the $W$ with zero bound
energy. Then $n$ eigenvalues can be obtained by the code of
DGEEV function in NAG Fortran Library.  If we are interesting in the
$s^{th}$ eigenvalues.  The eigenvalue with the fixed number $s$ is
substituted in the integral equation kernel on each iterative loop.
Then the linear spectral problem is solved again.  The stopping
criterion $|W^{(l)}_s-W_s^{(l-1)}|<\epsilon$ is tested on each
iteration. The value $\epsilon$ is chosen to satisfy the required
precision. As soon as stopping criterion is fulfilled, the iterative
process is terminated. The eigenvalue $W^{(l)}_s$ and eigenfunction
$\psi^{(l)}_s$ obtained on the last iteration are returned as the
solution of the problem.

\end{document}